\documentclass[aps,showpacs,showkeys,preprint]{revtex4}
\usepackage{epsfig,amssymb,bm,graphics}

\begin{document}


\title{\boldmath In-medium modification and decay asymmetry of $\omega$ mesons\\
in cold nuclear matter}


 \author{A.I.~Titov$^{a,b}$ and B.~K\"ampfer$^{a,c}$}
 \affiliation{
 $^a$Forschungzentrum Dresden-Rossendorf, 01314 Dresden, Germany\\
 $^b$Bogoliubov Laboratory of Theoretical Physics, JINR,
  Dubna 141980, Russia\\
 $^c$ Institut f\"ur Theoretische Physik, TU~Dresden, 01062 Dresden,
 Germany
 }


\begin{abstract}
 We discuss an asymmetry of the decay $\omega\to e^+e^-$
 in nuclear matter with respect to the electron and positron energies.
 This asymmetry is sensitive
 to the properties of the $\omega$ meson self-energy and, in particular,
 it has a  non-trivial dependence on the $\omega$ energy and momentum.
 Therefore,
 this asymmetry  may serve as a powerful tool in studying the properties
 of the $\omega$ meson in the nuclear medium.
 \end{abstract}

 \maketitle


\section{Introduction}

The study of in-medium properties of light vector mesons is a
topic of great importance for hadron and heavy-ion physics.
Besides the hope to get information on the mechanism of how
hadrons acquire their masses, the in-medium modifications are
related to the chiral symmetry restoration and changes of the QCD
vacuum structure \cite{RW2000}. Otherwise, vector mesons decaying into $e^+e^-$ pairs  can
provide unique information about the evolution of nuclear matter
in relativistic heavy-ion collisions as a penetrating
probe, not suffering final state interactions \cite{RW2000}.
Vector meson properties may not only be changed in compressed and
heated strongly interacting matter but also at normal baryon
density and zero temperature.

We will concentrate on the
properties of the $\omega$ meson  being a hadron with a small
decay width thus providing a unique probe for the expected in-medium modifications.
The interaction of the $\omega$ meson with surrounding nucleons, nucleon
resonances, and exchange-meson currents are thought to lead to
modifications of  $\omega$ meson propagating in a nucleus. First
experimental results of such modifications have been reported in
Refs.~\cite{Trnka05,Naruki06}. Much theoretical work has been devoted
for an evaluation of the $\omega$ meson self-energy as key quantity
for the prediction of spectral properties.
The calculations point to an increase of $\omega$ width reflecting
the opening of new inelastic channels. But the predictions for the $\omega$
spectral function, often condensed into one quantity -- the
mass, are wildly  differing in different models as the scale (and even the sign)
of the mass modification depends on the assumed dynamics of the $\omega$
interaction with the ambient nuclear medium. For instance, the models of
Refs.~\cite{Caillon95,KKW97,KWW99,Saito98,Dutt03} predict a shift of the
peak position of the $\omega$ spectral function to lower
energy, whereas results of some other
approaches \cite{Zschocke02,Post00,Lutz01,Dutt01,Steinmueller06,Mosel06}
show an upward shift.

As a rule, the available estimates of the self-energy are made
within the low-energy theorem \cite{Dover71}, where the
in-medium part of the self-energy is expressed through the $\omega
N$ forward scattering amplitude. In fact, in different models,
different components and channels of the $\omega N$ amplitude are
assumed to be important. For simplicity, in the following discussion
we limit our considerations to the contribution of such baryon
resonances which seem to be dominant not only in the Compton
$\omega N$ scattering amplitude but also in $\omega$ production in
photon-nucleon
\cite{Ajaka06,TL02,PennerMosel02,Shklyar05,Zhao01},
meson-nucleon \cite{PennerMosel02-,Lutz02} and
proton-proton ($p p$) reactions \cite{TN03,Fuchs02} near the threshold of
$\omega$.

In different models, different resonances become dominant. Thus, for example, for
$\omega$ meson production in $pp$ reactions, Ref.~\cite{TN03} argues
for a dominance of $P_{11}(1710)$ and $D_{13}(1700)$
resonances, while in Ref.~\cite{Fuchs02} the dominant contribution
comes from $S_{11}(1535)$ and $S_{11}(1650)$. A similar situation is met in
$\omega$ meson photo-production:
In Refs.~\cite{Lutz01,Lutz02}, a strong
contribution to the $\omega N$ channel comes from $S_{11}(1535)$,
$S_{11}(1650)$, and $D_{13}(1520)$ resonances, while in
Refs.~\cite{TL02,Shklyar05,Zhao01} a significant contribution stems from
spin-$5/2$ resonances. In Refs.~\cite{TL02,Zhao01,Shklyar05}, it is
$F_{15}(1680)$, while the analysis of Ref.~\cite{PennerMosel02}
supports a $P_{11}(1710)$ resonance. These ambiguities are
extended to the $\omega$ meson self-energy and the related
current-current correlation function.

In order to reduce the mentioned ambiguities, it would be nice to have,
together with the position and width of the resonance in the
current-current correlation function, additional observables
being sensitive to the dynamics of the $\omega N$ interaction. One
possible candidate is the asymmetry of the di-electron angular
or energy distribution related to the difference of the transverse
and longitudinal parts of the $\omega$ meson self-energy in a
nuclear medium. This difference disappears for the $\omega$ meson
at rest (relative to the nuclear medium) and becomes finite for a
finite $\omega$ meson momentum ${\bf q}$. High-spin resonances
can be excited only by the orbital interaction,
and therefore, they do not contribute at ${\bf q}=0$. When $|{\bf
q}|$ increases, they become important. The dependence of the
transverse and longitudinal parts of the partial amplitudes on the
$\omega$ meson mass and momentum and, therefore, the asymmetry
between transverse and longitudinal parts of the current-current
correlation function may be used as a tool for fixing the mechanism
of the relevant $\omega N$ interaction. This has a practical aspect since
most of experiments studying the in-medium $\omega$ meson
properties are dealing with non-zero $\omega$ meson momenta.

The aim of our paper is the discussion of an example of such an asymmetry.
We introduce the asymmetry between the transverse and longitudinal
parts of the current-current correlation function and, as mere
illustration, we apply it to a simple resonance model of the
$\omega N$ scattering. In fact, it is the same model which was
used previously for $\omega$ photo-production in Ref.~\cite{TL02}.

Our paper is organized as follows.
In Sec.~II we introduce the asymmetry of the $\omega\to e^+e^-$ decay.
In Sec.~III we present a model for current-current correlation function
in nuclear medium, where the $\omega N$ interaction is described
by a resonance model.
Numerical results and a discussion are given in Sec.~IV.
The summary can be found in Sec.~V.
The appendices contain a brief recollection of the $\omega$ self-energy in vacuum
and the effective Lagrangians and coupling strengths employed here.

\section{\boldmath Asymmetry of  $\omega\to e^+e^-$ decay}

We assume that the $\omega$ meson is produced in the inclusive reaction
$a + A \to \omega + X$,
where the projectile $a$ may be a photon or hadron.
Then, the $\omega$ meson propagates in the nuclear medium of the target nucleus $A$
and afterwards decays into a electron-positron pair.
Our starting point is the well known
relation between the differential di-electron production rate
$dR$ and the current-current correlator (CCC) $\Pi_{\mu \nu}$ \cite{Weldon90,GK91}
\begin{eqnarray}
\label{E1}
dR=\frac{e^2}{q^4}{\cal L}^{\mu\nu}\,{\rm Im}{\Pi}_{\mu\nu}
\frac{d{\bf p}_+}{2E_+}\,\frac{d{\bf p}_-}{2E_-} .
\end{eqnarray}
In the following, the four-momenta of
$\omega$ meson, outgoing electron and positron
are denoted as $q=(\omega, {\bf q})$ and
$p_\pm=(E_\pm, {\bf p}_\pm)$,
respectively. The masses of the virtual $\omega$ meson and electron/positron
are defined as $M=\sqrt{q^2}$ and $M_e$, respectively.
The quantization axis
$(\bf z)$ is chosen along direction of the $\omega$ meson velocity.
The lepton tensor reads
\begin{eqnarray}
\label{E2}
\frac14\, {\cal L}_{\mu\nu}=
{p_+}_\mu {p_-}_\nu +{p_-}_\mu {p_+}_\nu
- g_{\mu\nu}\,({p_-}\cdot{p_+} +M_e^2) ~.
\end{eqnarray}
The imaginary part of the CCC can
be decomposed in  longitudinal and transverse parts as
\begin{eqnarray}
\label{E3}
-{\rm Im}{\Pi}_{\mu\nu}= W^LP^L_{\mu\nu} + W^TP^T_{\mu\nu}~,
\end{eqnarray}
where $P^L_{\mu\nu}$ and $P^T_{\mu\nu}$
are the standard longitudinal and
transverse polarization tensors,
\begin{eqnarray}
\label{E4}
P^L_{\mu\nu}&=&
\left(u_\mu - q_\mu\frac{u\cdot q}{q^2}\right)
\left(u_\nu - q_\nu\frac{u\cdot q}{q^2}\right)
\frac{q^2}{q^2u^2 - (u\cdot q)^2}~,\nonumber\\
P^T_{\mu\nu}&=& P^0_{\mu\nu}-P^L_{\mu\nu}~,\qquad
P^0_{\mu\nu}=\left(g_{\mu\nu}-\frac{q_\mu q_\nu}{q^2}\right)~,
\end{eqnarray}
and $u=(1,{\bf 0})$ is the four-velocity of the medium wherein the
$\omega$ meson propagates. Then, one gets with $q=p_++p_-$ the
following relations
\begin{eqnarray}
\label{E5}
{\cal L}^{\mu\nu}\,P^L_{\mu\nu}&=&
-2q^2\left( 1- \frac{(E_+-E_-)^2}{{\bf q}^2} \right)~,\nonumber\\
{\cal L}^{\mu\nu}\,P^T_{\mu\nu}&=& -2q^2\left(1 +
\frac{(E_+-E_-)^2}{{\bf q}^2} + \epsilon \right)
\end{eqnarray}
with $\epsilon=4M_e^2/M^2$. We  are interested in di-electron
production with invariant mass $M\gg 2M_e$ and, therefore, all
terms proportional to $\epsilon$ can be omitted.

For later purposes we introduce a useful variable in the medium's rest frame
\begin{eqnarray}
 \label{E6}
\xi=\frac{E_+-E_-}{|{\bf q}|}~.
 \end{eqnarray}
 The electron energy $E_e$ and the variable $\xi$ are defined in the regions
\begin{eqnarray}
 \label{E7}
 \omega - |{\bf q}|
 \le 2E_e\le \omega + |{\bf q}|,
 \qquad
 -1\le\xi\le 1.
 \end{eqnarray}
 Integrating Eq.~(\ref{E1}) over the azimuthal angle of the decay plane,
 one can express the  differential di-electron production
 rate in terms of the variables $\xi$ and $q$
\begin{eqnarray}
 \label{E8}
 \frac{d^5R(M,{\bf q})}{d^4qd\xi}
 =\frac{\pi e^2}{2 q^2} \left( W^L(1-\xi^2) + W^T(1+\xi^2)
\right)~.
 \end{eqnarray}
Integration of this equation over $\xi$ leads to the famous rate
\begin{eqnarray}
 \label{E9}
\frac{d^4R(M,{\bf q})}{d^4q}
 =\frac{2\pi e^2}{3q^2}
\left( W^L + 2\,W^T \right)~.
 \end{eqnarray}

\begin{figure}[ht!]
\includegraphics[width=0.6\columnwidth]{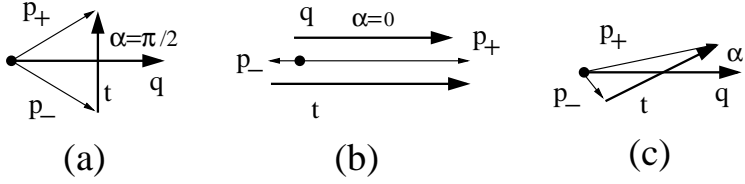}
\caption{\small{Geometry of the $\omega\to e^+e^-$ decay.
(a) and (b) correspond to cases where $\alpha=\pi/2$  (i.e. $\xi=0$)
and  $\alpha=0$ (${\rm i.e.}\,\, \xi=1$), respectively, while (c) depicts an intermediate
case  $0<\alpha<\pi/2$ (i.e. $0<|\xi\le 1$).
\label{Fig:1}}}
\end{figure}

Equation (\ref{E8}) allows us to define the asymmetry
\begin{eqnarray}
 \label{E10}
{\cal A}(\xi) = \frac{d^4R(\xi) - d^4R(0)}{d^4R(0)}=\xi^2\,{\cal
A}^{TL}
 \end{eqnarray}
with
\begin{eqnarray}
 \label{E11}
{\cal A}^{TL}
=
\frac{W^T - W^L}{W^T + W^L}~.
 \end{eqnarray}
One can see that the asymmetry with respect to the difference of
the electron and positron energies is directly related to the
asymmetry between longitudinal and transverse parts of the CCC.
The asymmetry can be
expressed as a function of the angle $\alpha$ between the two
vectors ${\bf q}={\bf p}_+ + {\bf p}_-$  and ${\bf t}={\bf p}_+ -
{\bf p}_-$,
 \begin{eqnarray}
 \label{E12}
\xi^2\simeq
\frac{\cos^2\alpha}{\cos^2\alpha +\gamma^2 \sin^2\alpha }~,
 \end{eqnarray}
where $\gamma=\omega/M$ is the Lorentz factor of the $\omega$
meson. The angle $\alpha$ is equal to $\pi/2$  for $\xi=0$ (i.e.
$E_+ =E_-$), and $\alpha=0$ for $\xi=0$ (i.e. $E_\pm =E_{\rm
{max\atop min}}$), as  depicted  in Fig.~\ref{Fig:1}.
Therefore, the asymmetry ${\cal A}(\xi)$ may be considered as
a measure of the anisotropy of the $\omega\to e^+e^-$
decay \cite{GK91,GTK96,TGK96}. For a slowly moving $\omega$ meson
with $\gamma\simeq 1$, we have $\xi^2\simeq \cos^2\alpha$, and the
angle $\alpha$ is close to the polar angle of the direction of the
electron momentum with respect to the direction of the $\omega$
meson momentum.

If the $\omega$ meson is in rest (${\bf q}=0$), with respect to
the medium's rest frame, then $W^T=W^L$ and the asymmetry ${\cal
A}^{TL}$ is equal to zero. At finite ${\bf q}$, the asymmetry
becomes non-zero and depends on the properties of the $\omega$ meson
self-energy~\cite{Weldon90,GK91,TGK96,GTK96}. In the general case,
${\cal A}^{TL}$ is a non-trivial function of $M$ and ${\bf q}$ as
shown below.

\section{Model for current-current correlation function in baryon medium}


Usually, the CCC is defined within the
vector dominance model based on the current-field
identity~\cite{Kroll67} for the electromagnetic current ${\cal J}_\mu$
\begin{eqnarray}
 \label{E13}
{\cal J}_\mu=-\sum\limits_V\frac{e M^2_V}{2\gamma_V}\,V_\mu~,
 \end{eqnarray}
where $M_V$ ($V=\rho,\omega,\phi...$) is the vector meson mass and
$2\gamma_V\equiv g_V$ is a dimensionless constant, which can be
determined from the partial $V\to e^+e^-$ decay widths
as $\gamma_\rho/ \gamma_\omega/ \gamma_\phi \simeq 2.5 / 8.5 / 6.7$.
We focus here on the $\omega$ contribution. An
explicit expression for the CCC can be derived using diagrammatic
technics (see, for example, Ref.~\cite{KKW97}).
The imaginary part of $\Pi$ is related to
the $\omega$ meson self-energy as
\begin{eqnarray}
 \label{E14}
\left[\frac{e^2M_{\omega 0}^4}{g_\omega^2}\right]^{-1}
{\rm Im}\,{\Pi^{cc}}^{L(T)}\equiv- W^{L(T)}
=
\frac{{\rm Im}\,{\Pi}_\omega^{L(T)}}
{|q^2 -M_{\omega 0}^2 - {\Pi}_\omega^{L(T)}|^2}~,
 \end{eqnarray}
where $M_{\omega 0}$ is a "bare" $\omega$ meson mass and
${\Pi_\omega}^{L(T)}$ represents the longitudinal (transverse)
part of the full $\omega$ meson self-energy. At low baryon
density, following the low-density theorem~\cite{Dover71}, the
full $\omega$ meson self-energy is expressed as a sum of its
vacuum part and  the contribution from the Compton $\omega N$
scattering \cite{KKW97,Mosel06}
\begin{eqnarray}
 \label{E15}
{\Pi_\omega}_{\mu\nu}= \Pi^{\rm vac}_{\omega\,\mu\nu}
+ \tilde\rho T^{\omega N}_{\mu\nu}~,
 \end{eqnarray}
where $\tilde\rho=\rho/2M_N$ and $\rho$ is the baryon density. The
nominator $2M_N$ arises from our choice of the Dirac spinor
normalization $\bar uu=2M_N$.  $T^{\omega N}_{\mu \nu}$ is the amplitude of
the Compton $\omega N$  scattering. For our purpose we decompose
it into longitudinal and transverse parts
\begin{eqnarray}
 \label{E16}
-T_{\mu\nu}=
T^LP^L_{\mu\nu} + T^TP^T_{\mu\nu}
 \end{eqnarray}
leading to
$ \Pi_\omega^{L(T)}=\Pi^{\rm vac}_\omega - \tilde\rho T^{L(T)}$
with
$\Pi^{\rm vac}_\omega = g^{\mu\nu} \,\Pi^{\rm
vac}_{\omega\,\mu\nu}/3$.
The vacuum self-energy determines the properties of $\omega$ meson
in vacuum, its decay width and the mass according to
\begin{eqnarray}
\label{E19}
M_\omega\Gamma_\omega &=&-{\rm Im}\,\Pi^{\rm vac}_\omega(q^2=M_\omega^2)~,
\nonumber\\
M_\omega^2&=&M_{\omega 0}^2+{\rm Re}\,\Pi^{\rm vac}_\omega(q^2=M_\omega^2)~,
\end{eqnarray}
and is discussed in many papers (cf.~\cite{KKW96,Mosel06}). Our choice,
based on the dominance of the $\omega\to\rho\pi$ transition
(Gell-Mann, Sharp, Wagner mechanism~\cite{GSW}), is close to that
of Ref.~\cite{Mosel06} and is described in Appendix~A. The
smoothly varying function ${\rm Re}\,\Pi^{\rm vac}_\omega$ is
absorbed in the mass parameter via $M_{\omega\,0} \to M_\omega$.

In general, the $\omega N$ scattering amplitude can be expressed
as a sum of two tensors
\begin{eqnarray}
\label{E20}
T_{\mu\nu}=a(p,q)P^0_{\mu\nu}+b(p,q)P^1_{\mu\nu}~,
\end{eqnarray}
where
\begin{eqnarray}
\label{E21}
P^1_{\mu\nu}=g_{\mu\nu} - \frac{p_\mu q_\nu + p_\nu q_\mu}{p\cdot q}
+\frac{p_\mu p_\nu q^2}{(p\cdot q)^2}~,
\end{eqnarray}
and $P^0_{\mu\nu}$ is defined in Eq.~(\ref{E4}). Then, using the relations
\begin{eqnarray}
\label{E22}
&&P^1_{\mu\nu}P^{T\mu\nu}=2,\qquad
P^0_{\mu\nu}P^{T\mu\nu}=2~,\nonumber\\
&& P^1_{\mu\nu}P^{L\mu\nu}=\frac{p^2q^2}{(p\cdot q)^2},\qquad
 P^0_{\mu\nu}P^{L\mu\nu}=1~,
\end{eqnarray}
one gets
\begin{eqnarray}
\label{E23}
T^{T}=a(p,q)+b(p,q),\qquad
T^{L}=a(p,q)+ \frac{p^2q^2}{(p\cdot q)^2}b(p,q)~.
\end{eqnarray}
The difference between the transversal and the longitudinal
parts of the $\omega$ meson self-energy is
\begin{eqnarray}
\label{E24}
\Pi^{T} -\Pi^{L}=
\frac{1-\gamma^2}{\gamma^2} b(p,q) =\frac{{\bf q}^2}{\omega^2}b(p,q) ~.
\end{eqnarray}
One can see that the absolute value of this difference increases
with the $\omega$ meson momentum and is quantified by the above
defined $\omega$ meson decay asymmetry.

\begin{figure}[ht!]
\includegraphics[width=0.4\columnwidth]{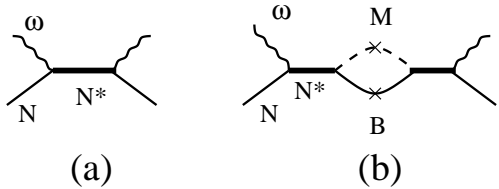}
\caption{\small{(a) Tree-level diagram for the Compton $\omega N$
scattering through nucleon resonance excitations.
(b) Schematic presentation of the imaginary part of the amplitude
 evaluated by cutting the intermediate baryon ($B$) and meson ($M$) lines.
\label{Fig:2}}}
\end{figure}

In order to illustrate the effect of such a difference we utilize a
resonance model for the $\omega N$ elastic scattering, depicted
schematically in Fig.~\ref{Fig:2}.

The amplitude is calculated as a sum of the tree-level $s$-channel
amplitudes shown in Fig.~\ref{Fig:2}~(a), where the imaginary
part is evaluated using the Cutkosky cutting rule. Thus, the contribution
of an individual resonance to the imaginary part of $T^{\mu\nu}$
is given as a sum of the cutted diagrams with intermediate two
particle states of $BM$ configurations shown in
Fig.~\ref{Fig:2}~(b) with $B=N,\Lambda,\Delta$ and
$M=\pi,\eta,K,\sigma\to2\pi,\rho\to 2\pi,\omega\to3\pi$
\begin{eqnarray}
\label{E25}
{\rm Im }T^{\mu\nu}_r=
\sum\limits_i\frac{p_i}{32\pi\sqrt{s}}
 \int {\rm Tr}[T^\mu_{r\,i} T^{\nu^*}_{r\,i}]\,d\cos\theta_i~.
\end{eqnarray}
Here $p_i$ and $\theta$ are the momentum of the intermediate meson of species $i$
and its polar angle in the $\omega N$ center of mass system, respectively;
$s$ is the square of the total energy.
$T^\alpha_{r\,\mu}\,\lambda_\alpha$, where $\lambda_\alpha$ is
polarization vector of the $\omega$ meson,
is the amplitude of the $\omega N \to N M$ transition via the intermediate
excitation of the resonance $N^*_r$.

Evaluating the real part of $T^{\mu\nu}$ we use the following assumption
\begin{eqnarray}
\label{E26}
{\rm Re }T^{\mu\nu}_r=
-{\rm Im }T^{\mu\nu}_r\frac{s-M_r^2}{M_r\Gamma_r}~,
\end{eqnarray}
where $M_r$ and $\Gamma_r$ are the mass and the total decay width
of the resonance $r$, respectively. The latter one depends on $s$. This
ansatz is motivated by the standard presentation of the resonant
amplitude as
\begin{eqnarray}
\label{E27}
T_r=\frac{A_r}{ s-M_r^2 + i M_r\Gamma_r}
\end{eqnarray}
which is widely used in similar analyses, e.g. in
Ref.~\cite{Eletsky01,Martell04}. Note that, if one could calculate the
resonant amplitude shown in Fig.~\ref{Fig:2}~(a), using the
standard Feynman rules, then Eq.~(\ref{E26}) would be exact. Our
analysis shows that for the most of important resonances the
imaginary part calculated either by Eq.~(\ref{E25}) or by a direct
calculation of the diagram shown in Fig.~\ref{Fig:2}~(a) are very
close to each other which encourage the use of the ansatz
Eq.~(\ref{E26}). Using Eqs.~(\ref{E25}) and (\ref{E26}), one can
derive the amplitudes $T^{TL}$ explicitly at finite momentum ${\bf q}$
and perform a quantitative analysis of the decay asymmetry. This
is the advantage of such a simplified model. The disadvantage
consists in a loss of the $S$ matrix unitarity due to lacking
coupling to additional transitions.  This is a typical
approximation for approaches based on tree-level diagrams.
Since the aim of the present paper is to show the qualitative
effect in the decay asymmetry, we leave a more detailed quantitative
analysis of the Compton amplitude (based, for example, on the
coupled channel approach of Ref.~\cite{Mosel06}) for future studies.

Evaluating the $\omega NN^*$ interaction we utilize the resonance
model previously used for the $\omega$-meson photo-production in
Ref.~\cite{TL02} within the effective Lagrangian formalism. We
consider isospin $I=1/2$ and spin $J\le 5/2$ nucleon
resonances listed in \cite{PDG} with the empirically known
helicity amplitudes of $\gamma N \rightarrow N^*$ transitions,
because the $\omega NN^*$ coupling constants are defined within the
vector dominance model. Exceptions are the $P_{11}(1710)$ and
$D_{13}(2080)$ resonances for which one can evaluate the $\omega
NN^*$ coupling constants from known partial decay widths for $N^*\to
\omega N$~\cite{PDG,Penner02}. We thus start our analysis by
taking into account contributions of the following 10
resonances: $P_{11}(1440)$,
$D_{13}(1520)$, $S_{11}(1535)$,
$S_{11}(1650)$, $D_{15}(1675)$,
$F_{15}(1680)$, $D_{13}(1700)$,
$P_{11}(1710)$,
$P_{13}(1720)$, and $D_{13}(2080)$. However, it turns out that only
six of them give a sizable contribution. These are
$D_{13}(1520)$, $S_{11}(1535)$,
$F_{15}(1680)$, $D_{13}(1700)$, $P_{11}(1710)$, and $D_{13}(2080)$.

 For the $N^{*}$ resonances with spin $J=1/2$, the effective Lagrangians for
 the $\omega NN^*$ interactions are chosen to be of the form of tensor coupling. This
 "minimal" form of Lagrangians, previously used in the study of
 $\eta$ photo-production \cite{BMZ95}, is
 \begin{eqnarray}
 {\cal L}_{\omega NN^*}^{\frac{1}{2}^{\pm}}
 =
 \,\, \frac{g_{\omega NN^*}}{2M_{N^*}} \bar\psi_{N^*} \,\Gamma^{(\pm)}
 \sigma_{\mu\nu} F^{\mu\nu} \psi_{N} +{\rm h.c.},
  \label{E28}
 \end{eqnarray}
 where $\psi_{N}$ and $\psi_{N^*}$ are the nucleon,
 and nucleon resonance fields, respectively, and
  $F^{\mu\nu}=\partial^\nu \omega^\mu - \partial^\mu \omega^\nu$,
 and $\omega_\mu$ is the $\omega$ meson field.
 The coupling
 $\Gamma^+ =1\,(\Gamma^-=\gamma_5$) defines the excitation
 of a positive (negative) parity $N^*$ state.

 For the $N^{*}$ resonances with spin $J = 3/2$,
 we use the expression introduced in
 Refs.~\cite{BMZ95,DMW91,OO75}
\begin{eqnarray}
 {\cal L}_{\omega NN^*}^{\frac{3}{2}^\pm}
  =
 i\frac{g^{}_{\omega NN^*}}{M_{N^*}}
 \bar\psi_{N^*}^\mu \,O_{\mu\nu}(Z)\gamma_\lambda\Gamma^{(\mp)}
 F^{\lambda\nu}
 {\psi_{N}} + {\rm h.c.},
 \label{E29}
 \end{eqnarray}
 where
 $\psi_\alpha$ is the Rarita-Schwinger  baryon field.
 The off-shell operator $O_{\mu\nu}(Z)$ is
 \begin{eqnarray}
 O_{\mu\nu}(Z)=g_{\mu\nu}-\left[
 \frac12 + Z \right]\gamma_\mu\gamma_\nu,
 \label{E30}
 \end{eqnarray}
 where $Z$ is the so-called "off-shell" parameter.

 The effective Lagrangians for the resonances with
 $J^P=\frac{5}{2}^{\pm}$ are constructed by analogy with the previous case,
\begin{eqnarray}
 {\cal L}_{\omega NN^*}^{\frac{5}{2}^\pm}
 =
 \frac{g^{}_{\omega NN^*}}{M_{N^*}^{2}}
 \bar\psi_{N^*}^{\mu\alpha} \,O_{\mu\nu}(Z)\gamma_\lambda\Gamma^{(\pm)}
 (\partial_\alpha F^{\lambda\nu})
 {\psi_{N}} + {\rm h.c.},
 \label{E31}
 \end{eqnarray}
 where  $\psi_{\alpha\beta}$
 is the spin-$5/2$ field.

The $\omega NN^*$ coupling is defined by using the vector dominance
model, which gives a relation between  $g^{}_{\omega NN^*}$ and the
iso-scalar electromagnetic coupling $eg^{s}_{\gamma NN^*}$
The determination of $eg^{s}_{\gamma NN^*}$ is described
in Ref.~\cite{TL02}. For the sake of convenience,
we list the employed coupling constants  $g^{}_{\omega NN^*}$ in Appendix~B.
There we also list the effective Lagrangians of the $N^* BM$ interaction
and the corresponding strength parameters.
Following \cite{TL02} we choose the off-shell parameter $Z=-1/2$
for all $N^*$.
Calculating the amplitude of the $\omega N\to N^*\to B M$ process
we parameterize the off-shell form factor of $N^*$ by the
covariant function
\begin{eqnarray}
F_{N^*}(p^2)=
\frac{\Lambda_{N^*}^4}{\Lambda_{N^*}^4  + (p^2-M_{N^*}^2)^2}~.
\label{E32}
\end{eqnarray}
The cut-off parameter $\Lambda_N=0.85$~GeV is taken to be the same
for all resonances.
The invariant amplitude of the transition $\omega N\to N^*\to N i$ has the
form of Eq.~(\ref{E27}), where the energy dependent total decay width
is calculated according to Ref.~\cite{Width},
\begin{eqnarray}
\Gamma^{\rm tot}(W)=\sum\limits_j
\Gamma_j\frac{\rho(W)}{\rho(M_{N^*})}~, \label{E32-}
\end{eqnarray}
where $\Gamma_j$ is the partial width for the resonance decay into
channel $j$, evaluated at $W=\sqrt{s}=M_{N^*}$. The form of the
"phase space-factor" $\rho(W)$ depends on the decay channel, the
relative momenta $k_j$ of the outgoing particles, and their
relative orbital momenta $l_j$. It provides the proper analytic
threshold behavior $\rho(W)\sim k_j^{2l_j+1}$ and becomes constant
at high energy \cite{Width2}.

In case of $J=3/2$ resonances we use the covariant modification of the
Rarita-Schinger propagator ${\cal P}_{\alpha\beta}$,
as suggested by Pascalutsa in \cite{Pascalutsa}.
For $J=5/2$ resonances, we use the covariant propagator
${\cal P}_{\alpha\beta,\delta\gamma}$ introduced in Ref.~\cite{Shklyar04}.

\section{ Results and Discussion}

First of all let us note that the $\omega N\to F_{15}$ transition comes
through the orbital interaction with $L=1,3$ and, therefore, the
resonance $F_{15}$ does not contribute to the $\omega$ meson
self-energy at $|{\bf q}|=0$. Formally, it follows from  the
identity
${\cal P}_{\alpha\beta,\gamma\delta}q^\alpha=0~, \label{E33} $
at ${\bf q}=0$.
Thus, at $|{\bf q}|=0$ the main contribution stems
from the excitation of the $D_{13}$  resonances. However, when
$|{\bf q}|$ increases, the contribution of the
$F_{15}$ resonance becomes important.

\begin{figure}[h!]
\includegraphics[width=0.4\columnwidth]{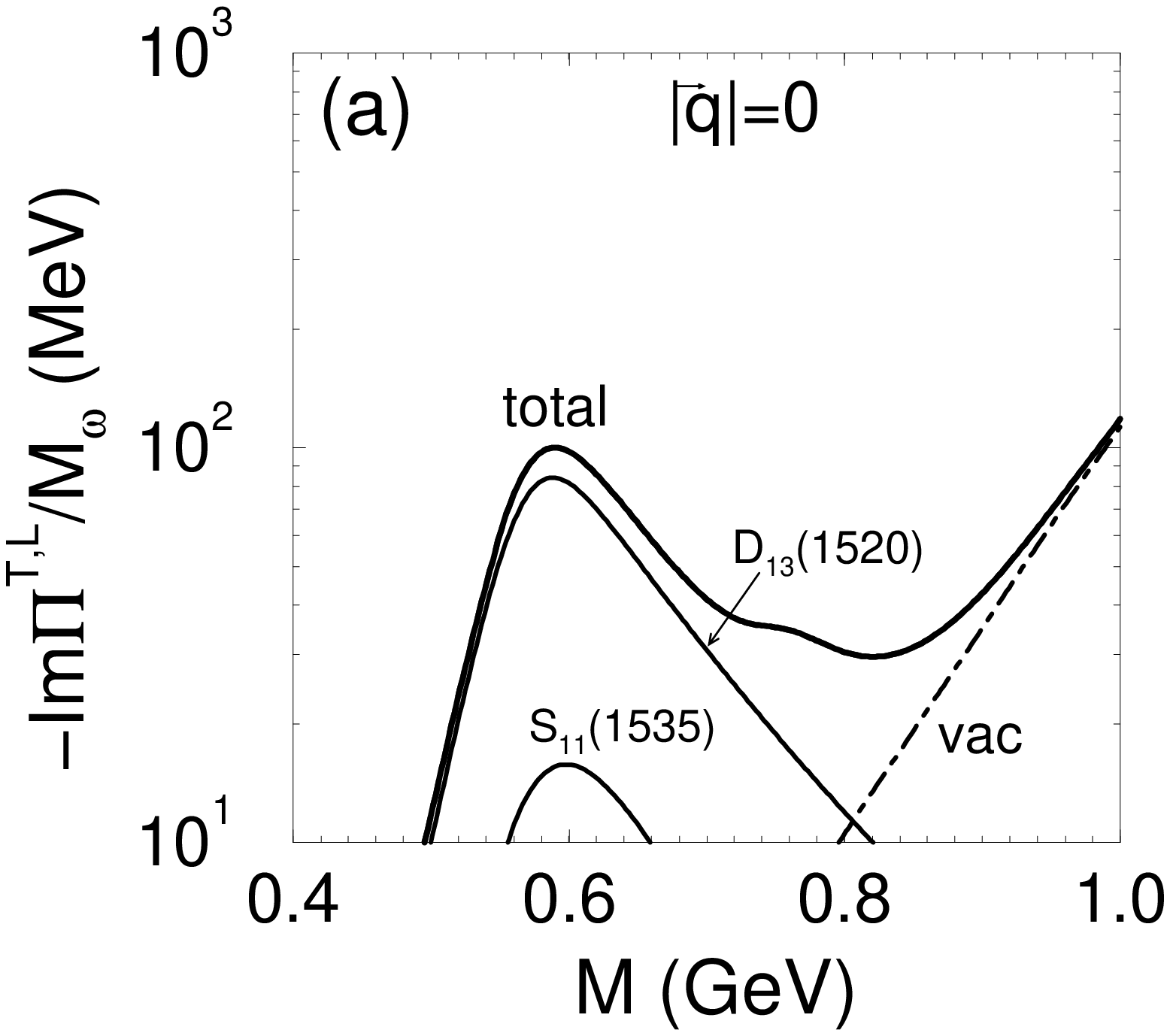}
\hspace{0.05\columnwidth}
\includegraphics[width=0.4\columnwidth]{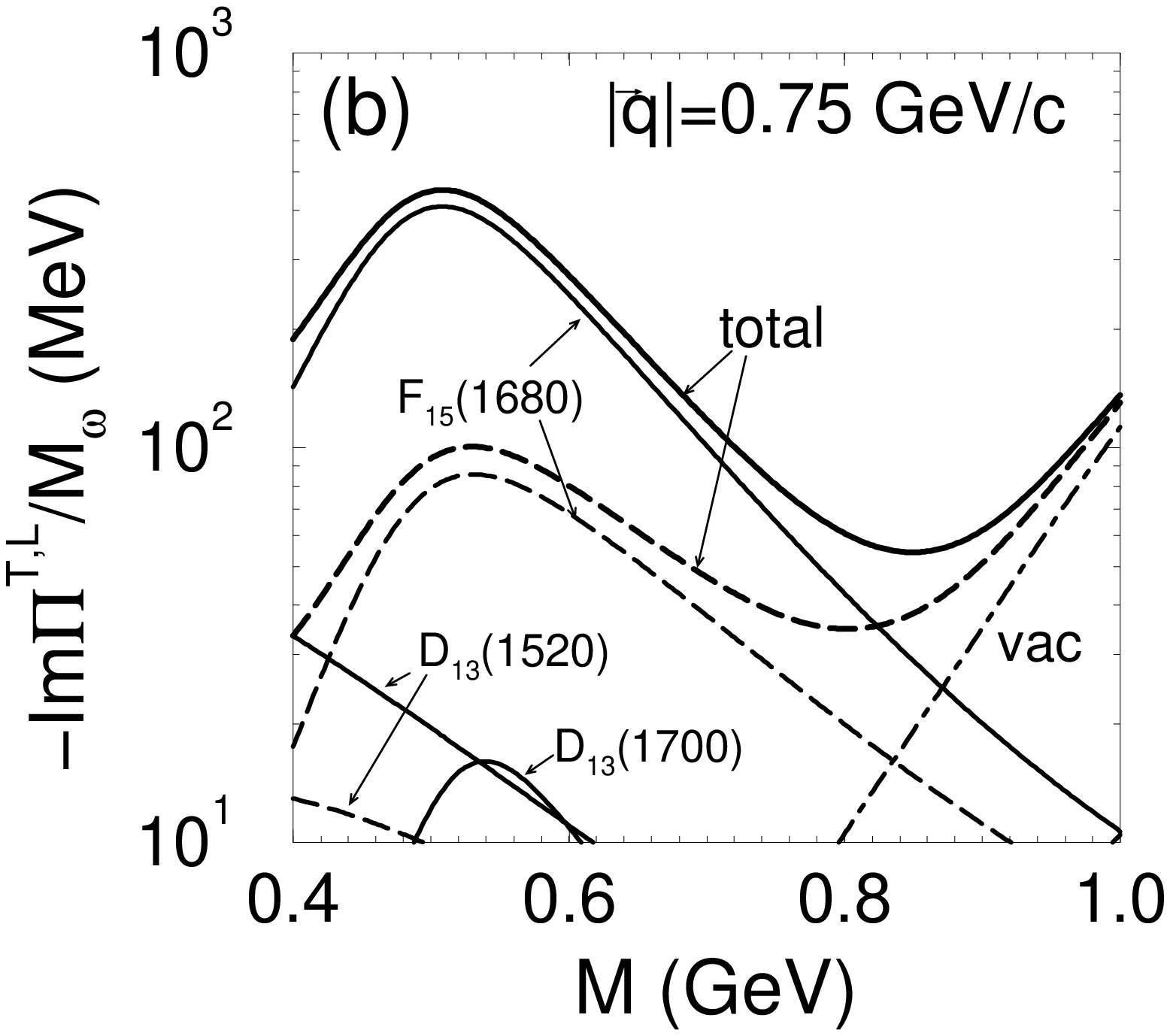}
\caption{\small{The imaginary part of the
$\omega$ meson self-energy at ${\bf q}=$ 0 (a) and 0.75 GeV/c
(b). The thin solid and dashed curves in (b)
correspond to the transverse and longitudinal components,
respectively. The thick solid curves correspond to the total
contributions. The vacuum contribution is depicted by dot-dashed curves.
\label{Fig:3} }}
\end{figure}

The imaginary part of the $\omega$ meson self-energy is presented
in Fig.~\ref{Fig:3}. We show the ratio $-{\rm
Im}\Pi^{TL}/M_\omega$ which approximately corresponds to the
modified $\omega$ meson decay width. The case of  $|{\bf q}|=0$ is
shown in Fig.~\ref{Fig:3}~(a). One can see that the main
contribution in our model comes, indeed, from the $D_{13} (1520)$
excitation. The next important resonance is $S_{11} (1535)$. The
contribution of other resonances is negligible. For $|{\bf
q}|=0.75$~GeV/c the situation changes. The dominant contribution
comes now from the $F_{15}$ resonance, while the $D_{13}$
excitations are much weaker. One can see a sizable difference
between transverse and longitudinal parts of the self-energy.

\begin{figure}[h!]
\includegraphics[width=0.4\columnwidth]{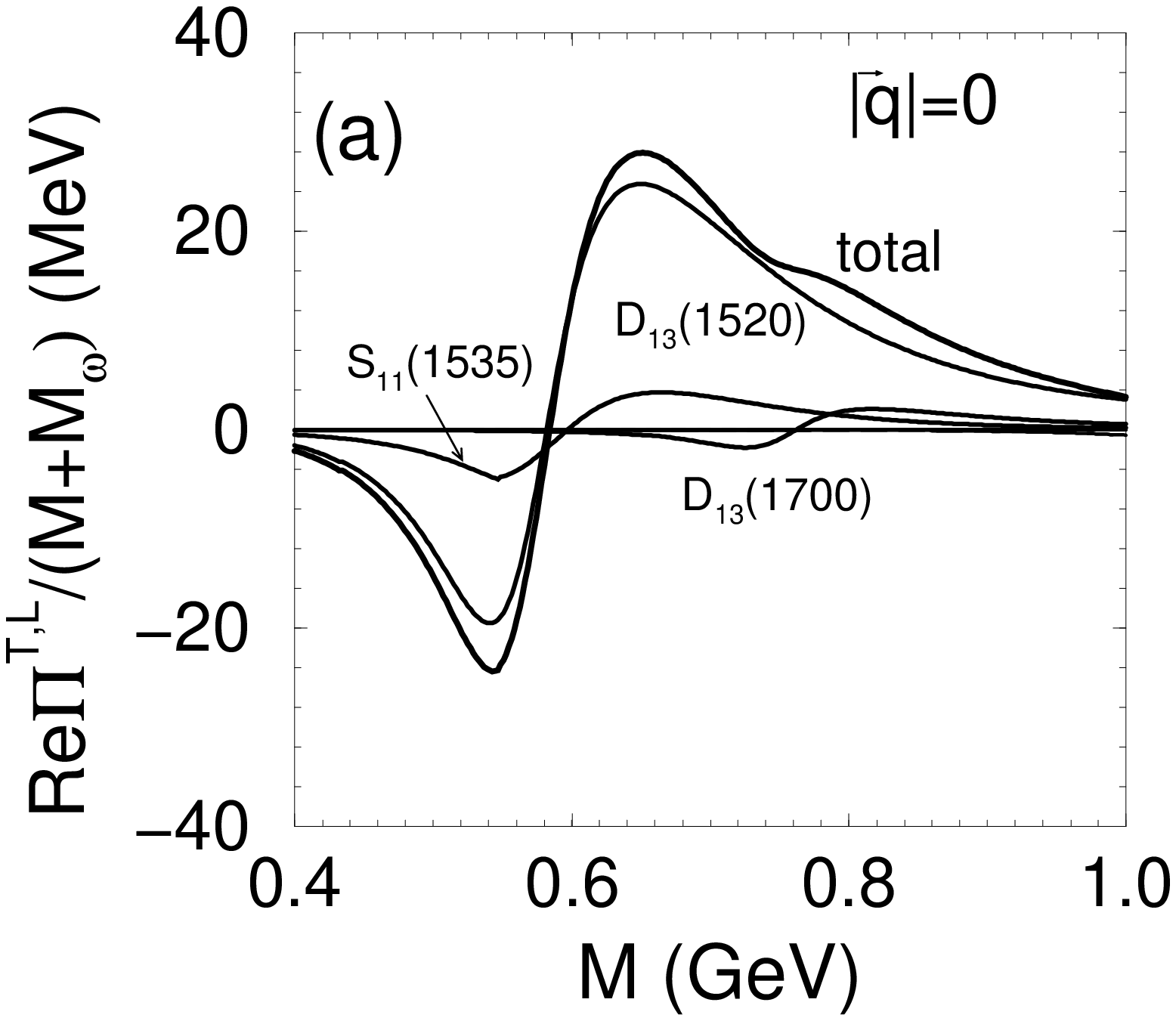}
\hspace{0.05\columnwidth}
\includegraphics[width=0.4\columnwidth]{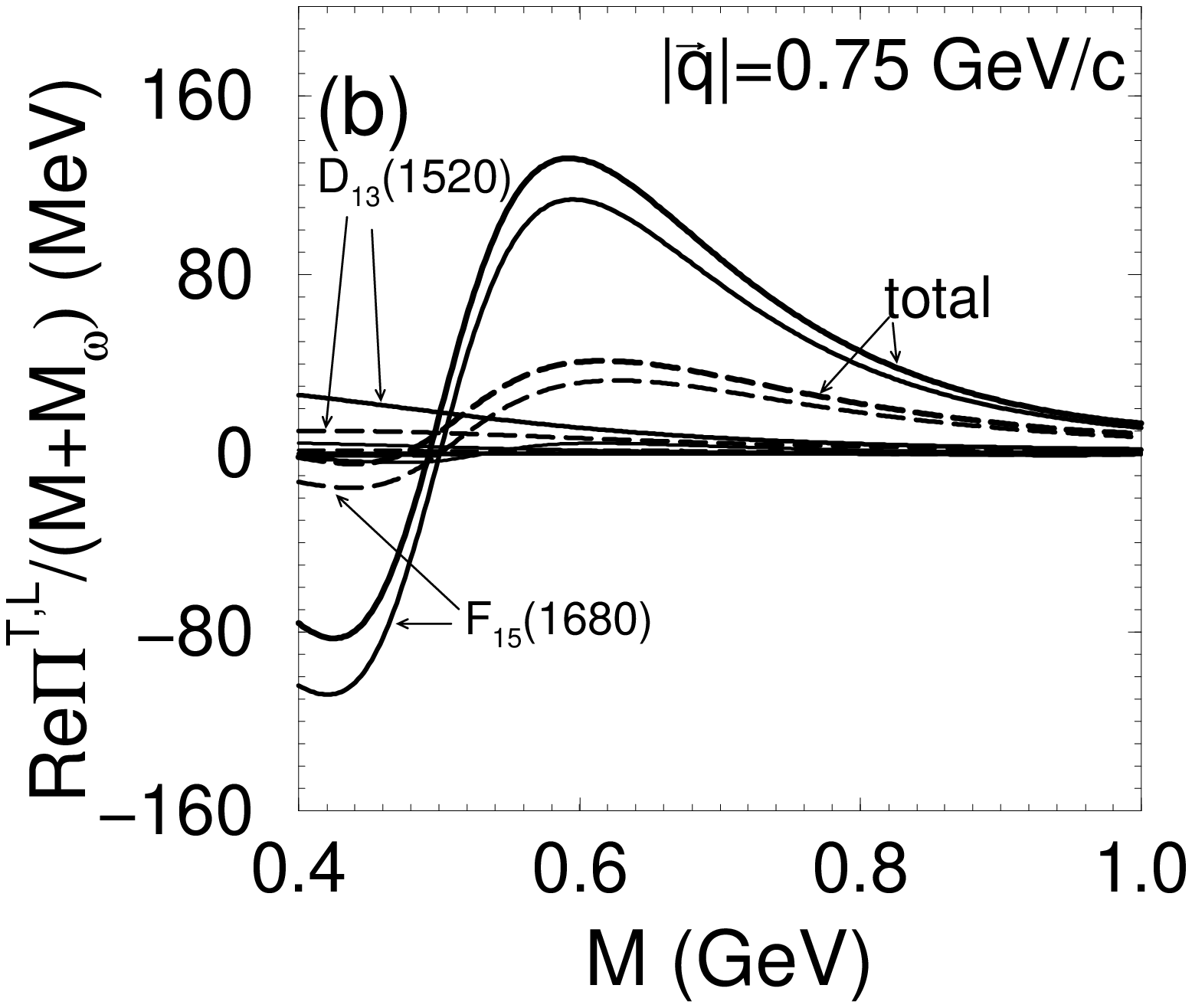}
\caption{\small{The real part of the
$\omega$ meson self-energy. Notation as in Fig.~\protect\ref{Fig:3}
\label{Fig:4} }}
\end{figure}

The real part of the self-energy is presented
in Fig.~\ref{Fig:4}. Here we show the ratio
${\rm Re}\Pi^{TL}/(M+M_\omega)$ which corresponds to
the shift of the  pole position of the
$\omega$ meson propagator in medium. Similar to the previous case,
for $|{\bf q}|=0$ the main contribution comes from
the $D_{13} (1520)$ excitation, while at $|{\bf q}|=0.75$~GeV/c
the $F_{15}$ resonance is dominant.

\begin{figure}[h!]
\includegraphics[width=0.4\columnwidth]{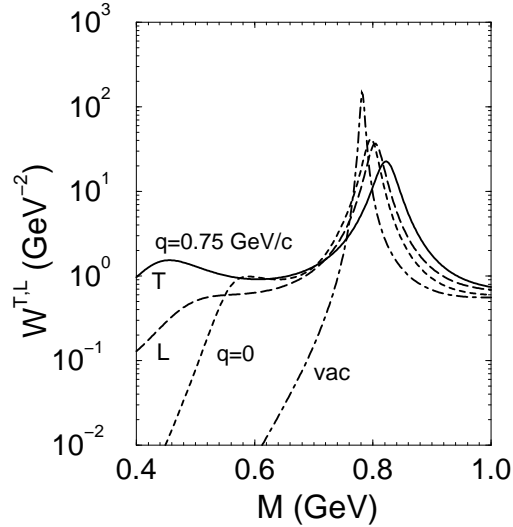}
\caption{\small{The functions $W^{T,L}$
for different values $|{\bf q}|$. The solid (dashed) curve depicts
$W^{T}$ ($W^{L}$). The vacuum case is shown by dot-dashed curve.
\label{Fig:5} }}
\end{figure}

The quantities $W^{T,L}$ as a function of the $\omega$ meson mass $M$
are presented in Fig.~\ref{Fig:5}. The vacuum case is shown by
the dot-dashed curve. It is evident that the model predicts some
(upward) sift of the pole position which increases from $\Delta
M=8$ to $20$ (40)~MeV for the longitudinal (transverse) part when
$|{\bf q}|$ changes from 0 to $0.75$~GeV/c.

\begin{figure}[h!]
\includegraphics[width=0.45\columnwidth]{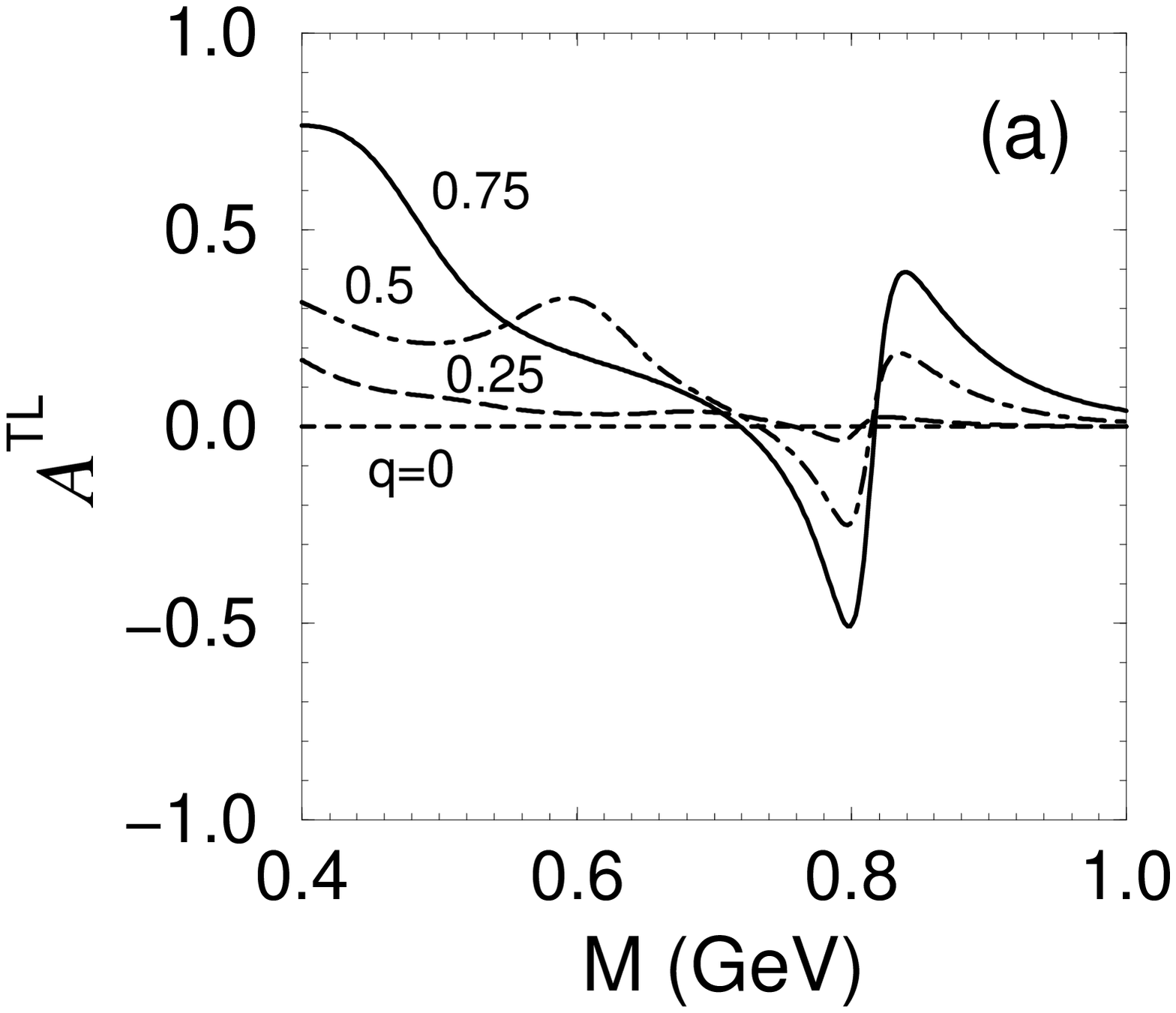}
\hspace{0.05\columnwidth}
\includegraphics[width=0.45\columnwidth]{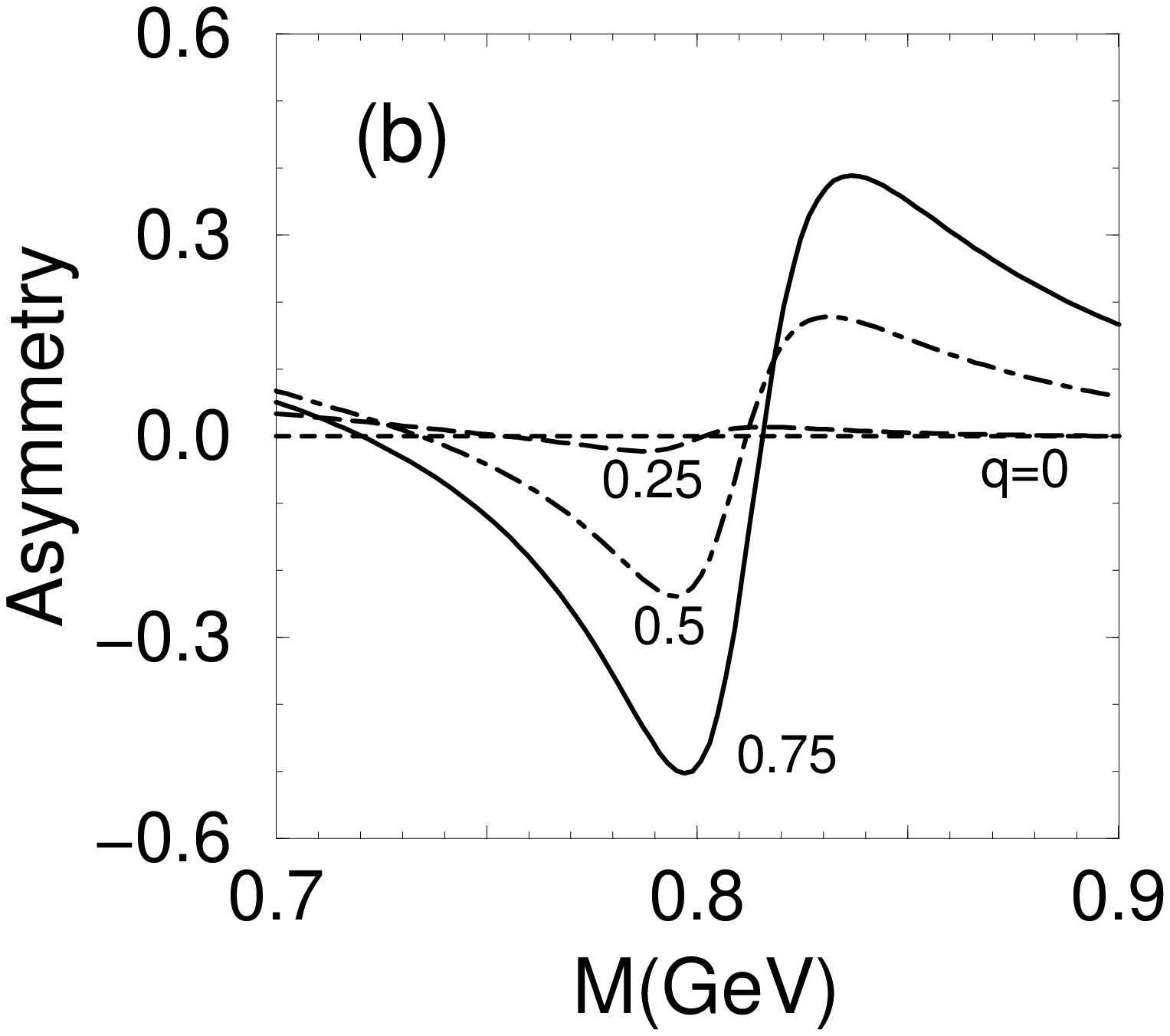}
\caption{\small{(a) The $\omega$ meson decay asymmetry
for different values $|{\bf q}|$. (b) The same as in (a) but at  $M\sim M_\omega$.
\label{Fig:6} }}
\end{figure}

Our result for the decay asymmetry is exhibited in Fig.~\ref{Fig:6}.
We display the asymmetry ${\cal A}^{TL}$ as a function of the
$\omega$ meson mass for different values of  $|{\bf q}|$. The asymmetry is
zero at $|{\bf q}|=0$. At finite $|{\bf q}|$ and $M\sim M_\omega$,
it becomes a non-trivial function of $M$ and it is defined by the
interplay of the different resonances, as seen in the left
panel of Fig.~\ref{Fig:6}. One can see that the sign and the
amplitude of the asymmetry depends strongly on both
$|{\bf q}|$ and $M$ and is sensitive to the details of the resonance model.
For $|{\bf q}| \simeq 0.75$~GeV/c the asymmetry is mainly determined
by the contribution of the  $F_{15}$ resonance. At $M \le 0.7$ it
is positive and monotonically increases with decreasing $M$. At
$M\sim M_\omega$ (see Fig.~\ref{Fig:6} (b)) its non-monotonic
behavior is determined by the difference of the peak positions for
$W^T$ and $W^L$. The shape of this curve again is sensitive to the
details of the resonance model.

\begin{figure}[ht!]
\includegraphics[width=0.3\columnwidth]{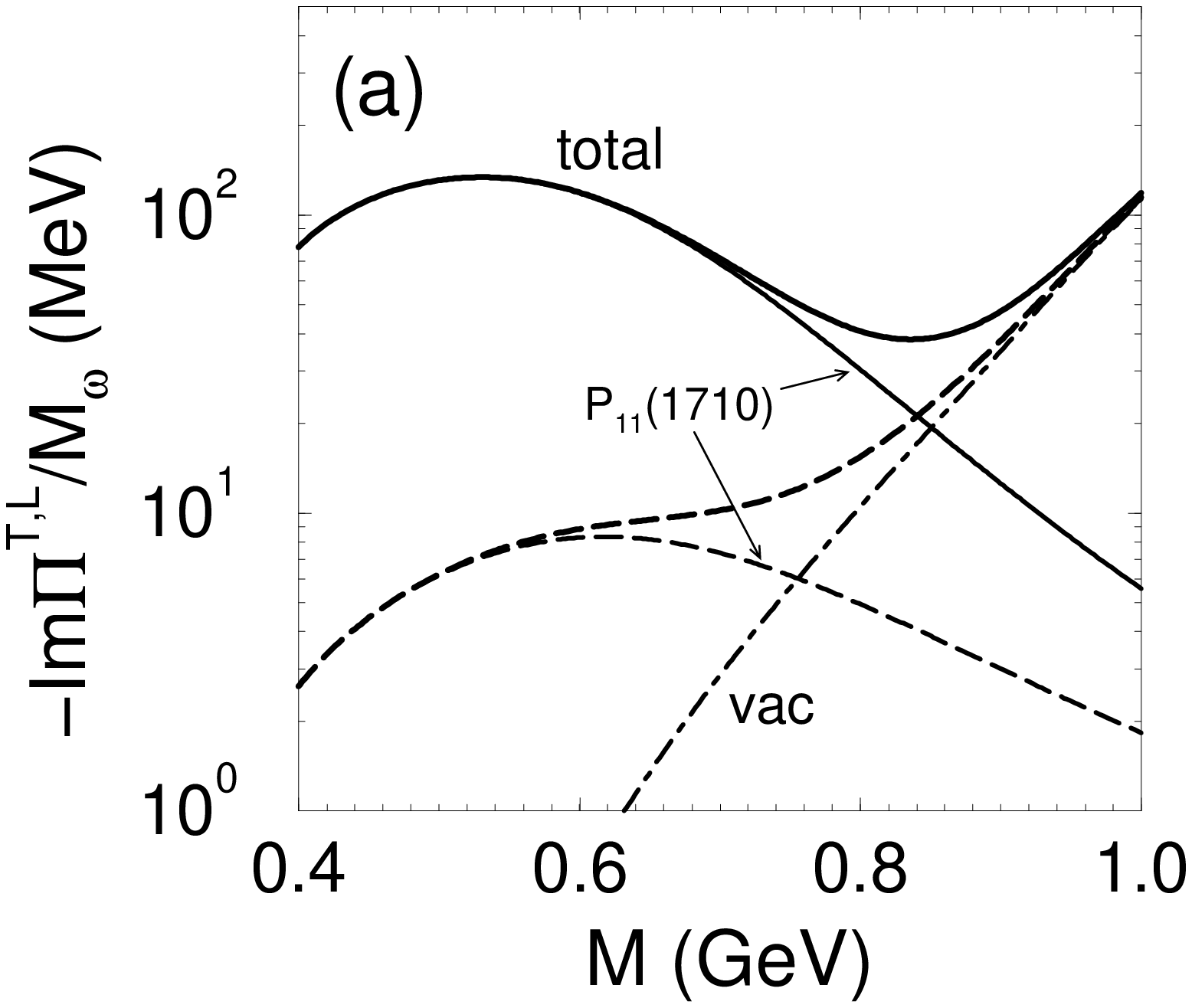}
\hspace{0.03\columnwidth}
\includegraphics[width=0.3\columnwidth]{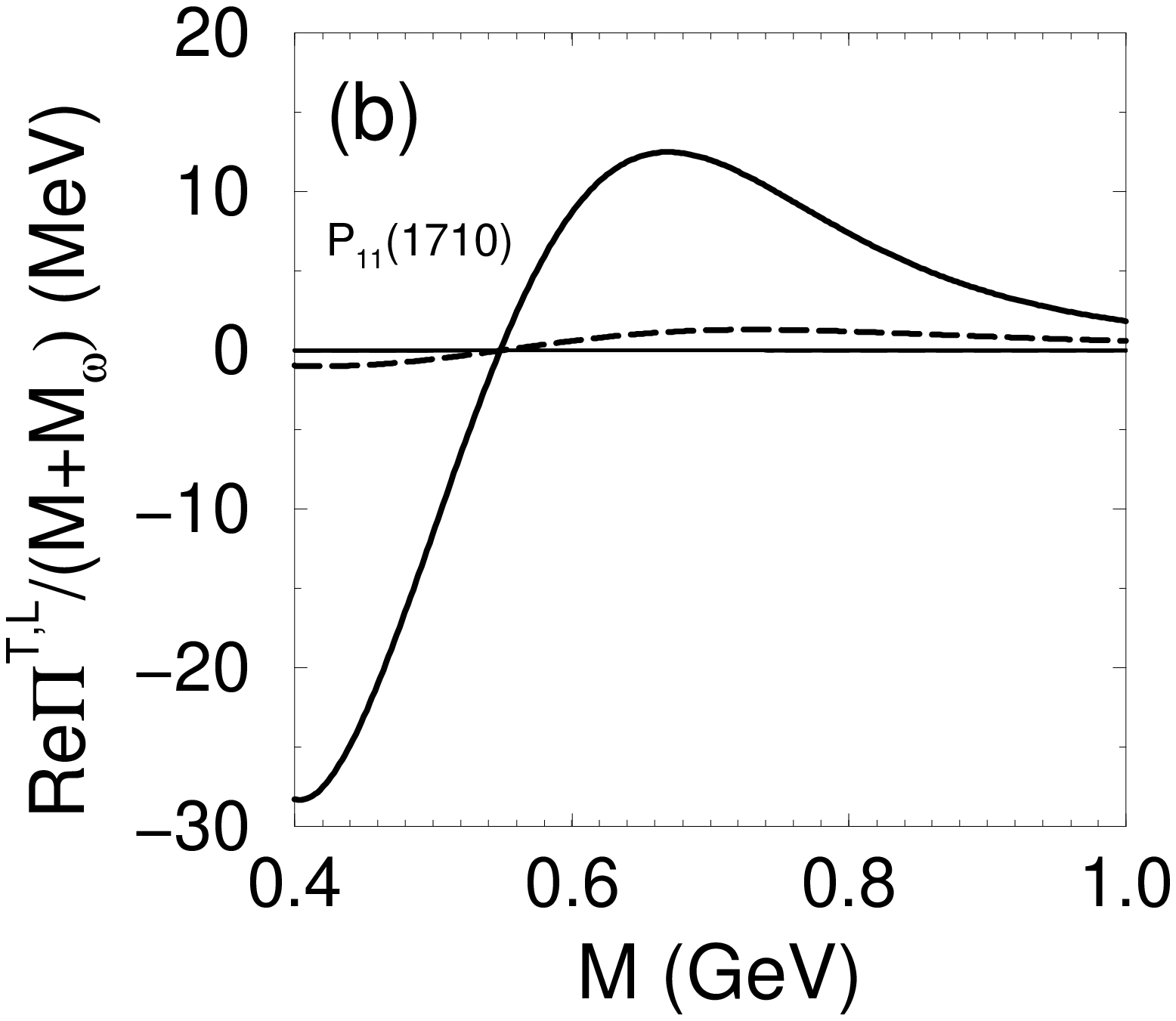}
\hspace{0.03\columnwidth}
\includegraphics[width=0.3\columnwidth]{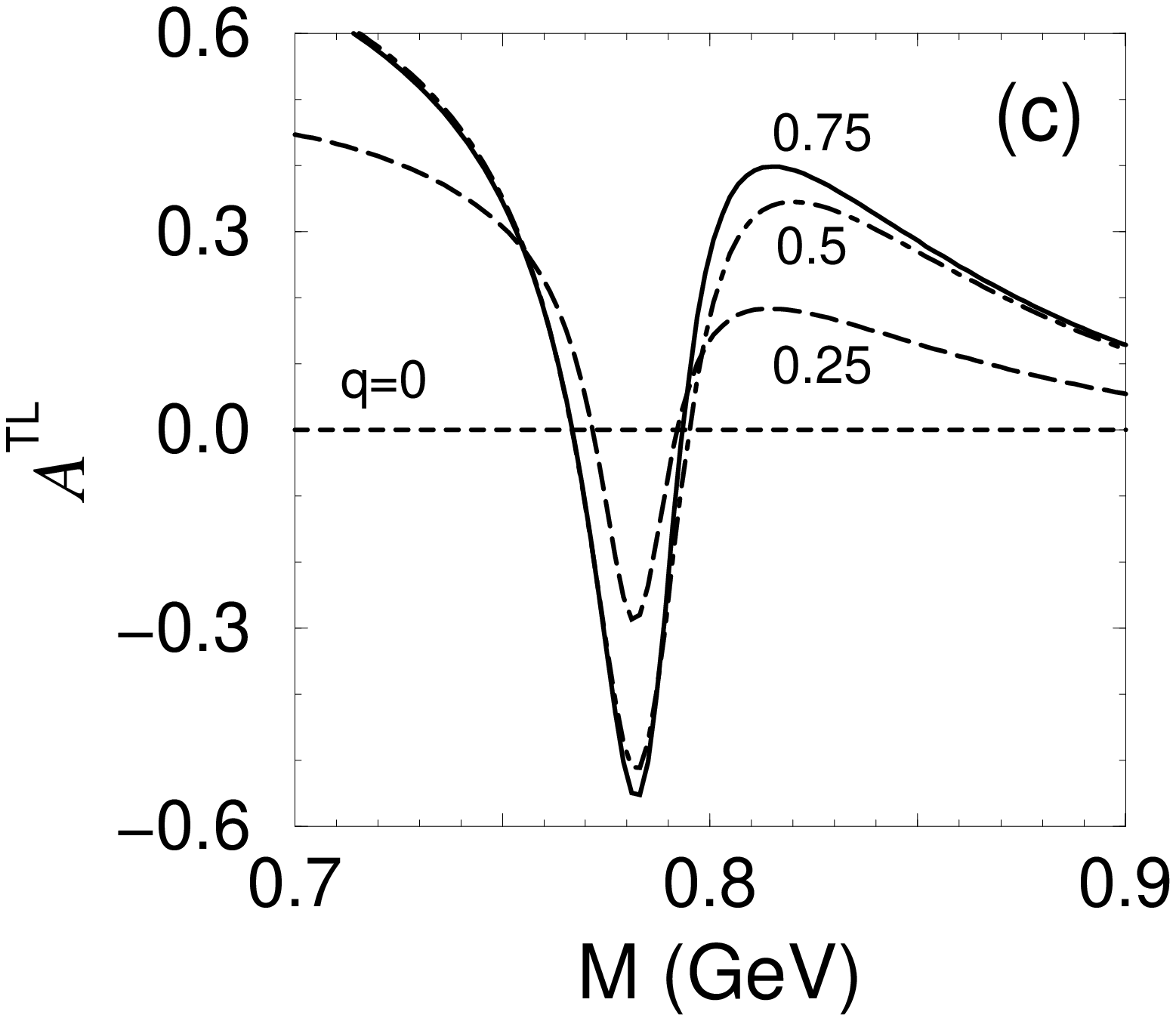}
\caption{\small{Results for the resonance model
dominated by the $P_{11}(1710)$ resonance.
(a) and (b) are for the imaginary and real parts of the $\omega$ meson
self-energy, respectively, while (c) exhibits the decay asymmetry.
\label{Fig:7} }}
\end{figure}

Note that our result for $\Pi^{T,L}_\omega$ in medium is
qualitatively similar to that of Ref.~\cite{Mosel06} obtained within a
coupled channel approach. Therefore, we expect that the asymmetry
for the model \cite{Mosel06} would be also close to our result.

\begin{figure}[h!]
\includegraphics[width=0.4\columnwidth]{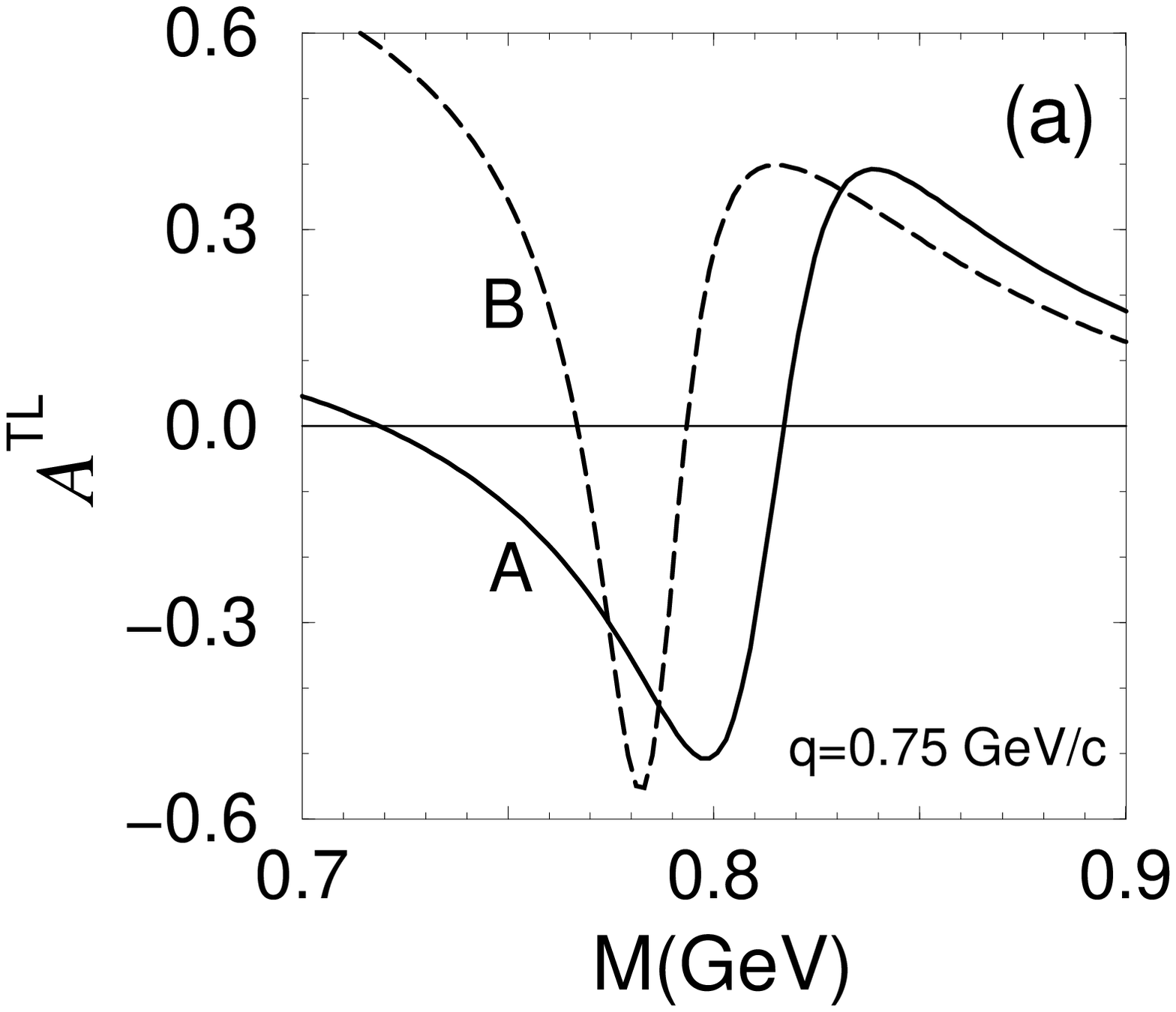}
\hspace{0.05\columnwidth}
\includegraphics[width=0.4\columnwidth]{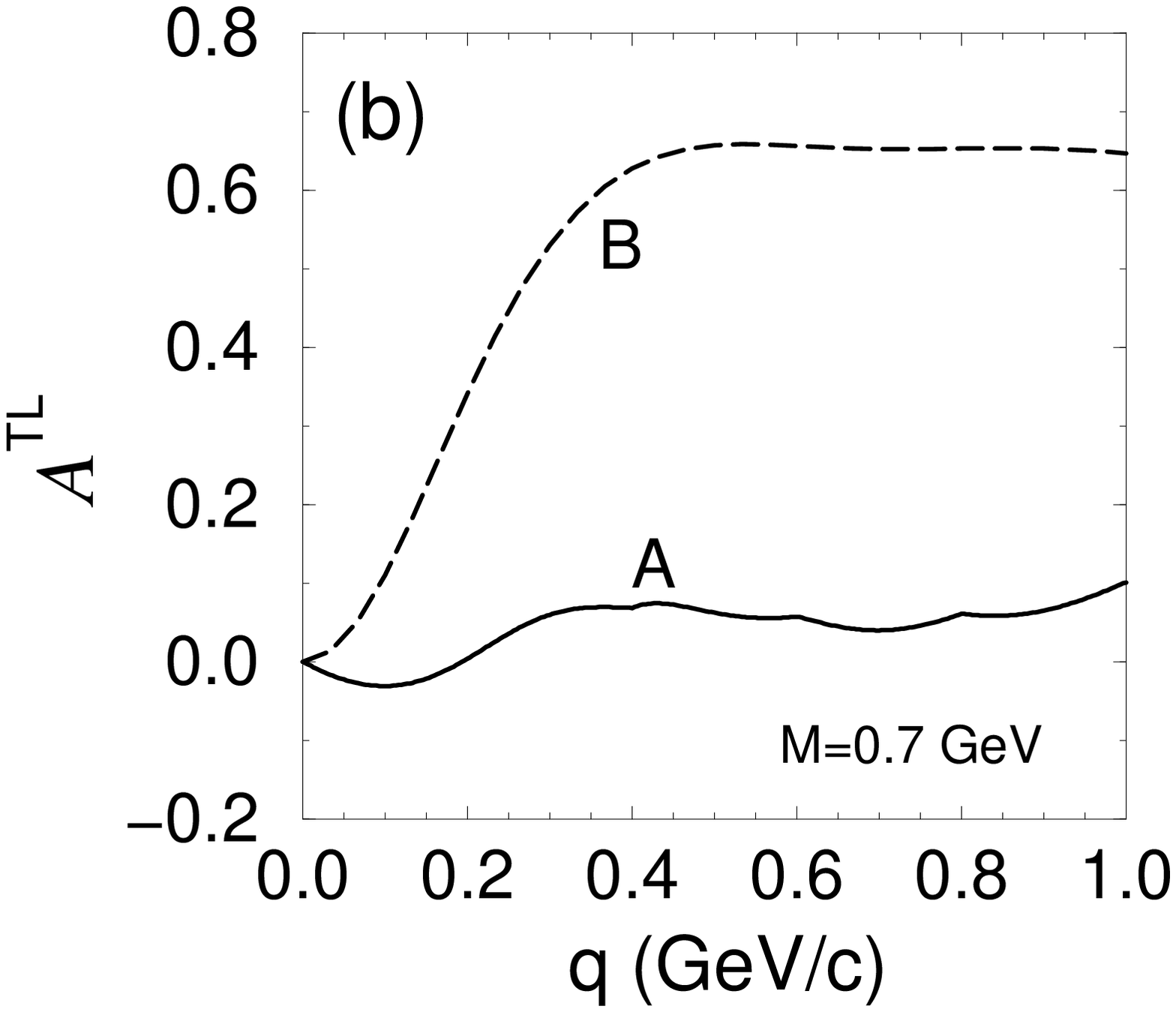}
\caption{\small{The $\omega$ meson decay asymmetry for models A
and B depicted by solid and dashed curves. (a) dependence on $M$
at $|{\bf q}|=0.75$~GeV/c, (b) dependence on $|{\bf q}|$ at
$M=0.7$~GeV. \label{Fig:8} }}
\end{figure}

In order to
demonstrate the sensitivity of the asymmetry to the actually employed
resonance model we show in Fig.~\ref{Fig:7}
results for the resonance model developed in
Ref.~\cite{TN03} for the reaction $pp \to \omega pp$. Here, the dominant
contribution comes from the $P_{11}(1710)$ resonance with the
$\omega NN^*$ coupling constant, rewritten in our notation,
$g_{\omega NN^*} = 10.32$, i.e.\ 4.9 times larger than ours.
The dependence
of the polarization operators on  the momentum ${\bf q}$ is
different, and this difference is manifest in the asymmetry. Thus,
there is no dramatic change of the asymmetry if $|{\bf q}|$
increases from 0.25 to 0.75~GeV/c, as seen in
Fig.~\ref{Fig:7}. To illustrate the difference between the two
models, we exhibit in Fig.~\ref{Fig:8} simultaneously the asymmetry
for these two models: the resonance model developed in Sec.~III (curves labelled by A) and
the model \cite{TN03}, dominated by the  $P_{11}(1710)$ resonance
(curves labelled by B). One can
see a strong difference in the dependence on $M$ at fixed $|{\bf
q}|=0.75$~GeV/c. Also the dependence on $|{\bf q}|$ at fixed
$M=0.7$~GeV is rather different. Such a difference in the
$M$ vs. $|{\bf q}|$ dependence may be used as a tool for studying the nature of
the $\omega$ meson self-energy.

\section{Summary}

In summary we discuss an asymmetry of the  $\omega\to e^+e^-$ decay
with respect to the electron and positron energies in
the nuclear-matter rest-system.
Thereby, we suppose that the $\omega$ meson is created by an
elementary projectile impinging on a heavy target nucleus.
The asymmetry is zero for the $\omega$ meson at rest and it is non-zero
at non-zero $\omega$ momentum. We find that the asymmetry is sensitive
to the properties of the $\omega$ meson self-energy and, in particular,
it has a non-trivial dependence on the $\omega$ mass (energy) and momentum.
We have shown that the excitation of high-spin
resonances  results in a strong momentum dependence of the
asymmetry around $M=0.75-0.8$~GeV and it is flat at  $M=0.7$~GeV.
Therefore, the asymmetry may be used as a powerful tool in
studying the properties of the $\omega$ meson in a nuclear medium.
Our analysis is performed by using a simple resonance model where
the coupling between $\omega N$ and $M B$ ($M = \pi, \eta, \sigma,
\rho, K;\,B = N, \Delta, \Lambda$) channels is taking into account
only in evaluating the imaginary part of the Compton $\omega N$
elastic scattering amplitude. An interesting subject for further
investigations is to study this asymmetry within more
sophisticated models, say within a consistent description of
coupled channels as well as on the basis of the QCD inspired
models. Experimentally, the asymmetry can be studied at various
facilities, e.g., KEK, HADES at GSI etc.

\acknowledgments

 We would like to thank V. Shklyar for fruitful discussions.
 One of the authors (A.I.T.) appreciates the FZD for the hospitality.
 This work was supported by BMBF grant 06DR136 and GSI-FE.

\begin{appendix}

\section{\boldmath $\omega$ meson self-energy in vacuum}

\begin{figure}[hb!]
\parbox{.3\textwidth}{\includegraphics[width=0.2\columnwidth]{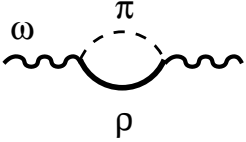}}
\hfill
\parbox{.6\textwidth}{
\caption{\small{The $\omega$ meson self-energy in vacuum.
\label{Fig:B1}}}}
\end{figure}

We assume the dominance of the virtual $\omega\to \rho\pi$ vacuum
transitions, depicted in Fig.~\ref{Fig:B1}.
This transition is described by the effective
Lagrangian in obvious standard notations
\begin{eqnarray}
\label{B1}
{\cal L}=\frac{g_{\omega\rho\pi}}{M_\omega}\,
\epsilon^{\mu\nu\alpha\beta}\partial_\mu\omega_\alpha
\partial_\nu\rho_\beta~.
\end{eqnarray}
The imaginary part of $\Pi^{\rm vac}$ is calculated, using Cutkosky rules, as
\begin{eqnarray}
\label{B2}
{\rm Im}\, \Pi^{\rm vac}_\omega(q^2)
=-\frac{g_{\omega\rho\pi}^2\sqrt{q^2}}{12\pi M_\omega^2}
\,p^3(q^2,M_\rho^2)~,
\end{eqnarray}
with
$p^2(q^2,M_\rho^2) = \lambda(q^2,M_\rho^2,M_\pi^2 )/( 4q^2)$.
Taking into account the $\rho$ meson mass distribution in the
region $2M_\pi<M_\rho<\sqrt{q^2}-M_\pi$ and normalizing  ${\rm
Im}\,\Pi^{\rm vac}_\omega(M_\omega^2)$ to the total decay width
$\Gamma_\omega$ we get
\begin{eqnarray}
\label{B4}
{\rm Im}\, \Pi^{\rm vac}_\omega(q^2) =-\sqrt{q^2}\,\Gamma_\omega
\frac{G(\sqrt{q^2})}{G(M_\omega)}
\end{eqnarray}
with
\begin{eqnarray}
\label{B5}
G(x)=\int\limits_{2M_\pi}^{x-M_\pi}
\frac{p^3(x^2,y^2)ydy}{(y^2 - M_\rho^2)^2 + (M_\rho\Gamma_\rho)^2}~,
\end{eqnarray}
where
$\Gamma_\rho=149.2$~MeV is the total $\rho$ meson decay width.

The real part of the self-energy depends on the regularization
scheme. Its absolute value is comparable, in order of magnitude,
with the absolute value of the imaginary part being much smaller
than the square of the physical $\omega$ meson mass. Therefore, to
avoid unknown parameters, we simply put the combination $M_{\omega
0}^2 + {\rm Im}\, \Pi^{vac}_\omega(q^2) \simeq M_{\omega 0}^2 +
{\rm Im}\, \Pi^{vac}_\omega(M_\omega^2)$ in denominator of
Eq.~(\ref{E16}) to be equal $M_\omega^2$, with replacing
$M_{\omega 0}^4\to M_{\omega}^4$ in the numerator.

\section{\boldmath Fixing parameters of the resonance model}

\subsection{\boldmath Effective Lagrangians for the $N^*\to \mu B$ transitions}

Consider first
the virtual $N^*\to \mu N $ transitions with $\mu=\pi,\eta,\sigma,\rho, \omega$.
The effective Lagrangian for the $\rho NN^*$ interaction is
taken to be the same as for $\omega NN^*$ (cf. Eqs.~(\ref{E28}) - (\ref{E31}))
with obvious generalization for the isospin $I_\rho=1$).
The interactions written in standard notation read
\begin{eqnarray}
 {\cal L}_{\mu NN^*}^{\frac{1}{2}^{\pm}}
 =
\bar \psi_{N^*}\left[
 g_{\pi NN^*}\Gamma^\mp{\bm{\pi\cdot\tau}} + g_{\eta NN^*}\Gamma^\mp \eta
+ g_{\sigma NN^*}\Gamma^\pm \sigma
\right]\psi_{N} +{\rm h.c.}~,
\label{C1}
 \end{eqnarray}
\begin{eqnarray}
 {\cal L}_{\mu NN^*}^{\frac{3}{2}^{\pm}}
 =
\bar \psi_{N^*}^\alpha\left[
\frac{g_{\pi NN^*}}{M_{N^+}}
\Gamma^\pm\partial_\alpha{\bm{\pi}\bm{\cdot\tau}}
+ \frac{g_{\eta NN^*}}{M_{N^+}}\Gamma^\pm \partial_\alpha\eta
+ \frac{g_{\sigma NN^*}}{M_{N^+}}\Gamma^\mp \partial_\alpha\sigma
\right]\psi_{N} +{\rm h.c.}~,
\label{C2}
 \end{eqnarray}
\begin{eqnarray}
 {\cal L}_{\mu NN^*}^{\frac{5}{2}^{\pm}}
 =i
\bar \psi_{N^*}^{\alpha\beta}\left[
\frac{g_{\pi NN^*}}{M_{N^+}}
\Gamma^\mp\partial_\alpha\partial_\beta{\bm{\pi}\bm{\cdot\tau}}
+ \frac{g_{\eta NN^*}}{M_{N^+}}\Gamma^\mp \partial_\alpha\partial_\beta\eta
+ \frac{g_{\sigma NN^*}}{M_{N^+}}\Gamma^\pm \partial_\alpha\partial_\beta\sigma
\right]\psi_{N} +{\rm h.c.}~,
\label{C3}
 \end{eqnarray}
where $\Gamma^+=1$ and  $\Gamma^-=\gamma_5$.
The interactions $N^*\to K\Lambda $ are similar to $N^*\to \eta N\Lambda $ with
substitutions $\eta\to K$ and $N\to \Lambda$.
The interactions $N^*\to \pi\Delta $ are chosen as
\begin{eqnarray}
 {\cal L}_{\pi \Delta N^*}^{\frac{1}{2}^{\pm}}
 =i \frac{g_{pi \Delta N^*}}{N^*}
\bar \psi_{N^*}\Gamma^\pm\partial_\alpha K
\psi_{\Delta}^\alpha + {\rm h.c.}~,
\label{C4}
 \end{eqnarray}
\begin{eqnarray}
 {\cal L}_{\pi \Delta N^*}^{\frac{3}{2}^{\pm}}
 = g_{\pi \Delta N^*} \bar \psi_{N^*}^\alpha\Gamma^\mp
\psi_{\Delta\,\alpha}K + {\rm h.c.}~,
\label{C5}
 \end{eqnarray}
\begin{eqnarray}
 {\cal L}_{\pi \Delta N^*}^{\frac{5}{2}^{\pm}}
 =i\frac{g_{\pi \Delta N^*}}{N^*}
\bar \psi_{N^*}^{\alpha\beta}\Gamma^\pm\partial_\beta K
\psi_{\Delta\,\alpha} + {\rm h.c.}~.
\label{C6}
 \end{eqnarray}

\subsection{\boldmath $N^*BM$ coupling constants}

In Table~I we present the coupling constants of $N^* \to B M$
transitions for the dominant amplitudes.
\begin{table}[h!]
\caption{The coupling constants of the employed effective Lagrangians.
The $g_\omega NN^*$ coupling constants are from Ref.~\cite{TL02}
(Model~II) using the vector dominance model. For $P_{11}(1710)$
and $D_{13}(2080)$ it is deduced from the branching ratios of
$N^*\to N\omega$ decays being 13\% and 21\% \cite{PDG}.
Other coupling constants are found by fitting the corresponding
branching ratios of the $N^*\to N\mu$ decay taken from
Ref.~\cite{PDG} and shown in parentheses (in \%). We show only
absolute values of the coupling constant, as their phases
drop out in our calculations.} \label{tab:I}
\begin{ruledtabular}
\begin{tabular}{lllrrrrrrrr}
$N^*$& $M_{N^*}$ & $g_{\omega NN^*}$& $g_{\pi NN^*}$ & $g_{\eta NN^*}$ & $g_{\sigma NN^*} $ &$g_{\rho NN^*}$ &$g_{K\Lambda N^*}$ &$g_{\pi \Delta N^*}$
&  \\ \hline
$S_{11}$ & $1535$ & $2.14$& $0.684(45)$  & $2.12(55)$ & $ - $ &$ - $ &$ - $ &$ - $ &\\
$P_{11}$ & $1710$ & $2.12(13)$& $ 2.28(15)$  & $4.22(6.2)$ & $1.51(25)$ &$1.13(15)$&$13.4(15)$ &$1.66(10.8)$ &\\
$D_{13}$ & $1520$ & $5.70 $   & $17.9(60)$ & $ -$ & $-$ &$1.40(20) $ & $- $ &$0.365(20)$ & \\
$D_{13}$ & $1700$ & $1.16$   & $5.36(15)$ & $ - $  &$6.77(82)$ &$ -$ &$4.97(3)$ & $-$ &  \\
$D_{13}$ & $2080$ & $2.91(21)$   & $8.27(23)$ &  $ 11.0(7)$ & $ -$ & $-$ &$-$ &$0.708(49)$ &  \\
$F_{15}$ & $1680$ & $35.0$    & $62.0(65)$ & $ -$ & $ 21.1(25)$ &$-$ & $-$ & $1.72(10)$ & \\
\end{tabular}
\end{ruledtabular}
\end{table}

\end{appendix}

\end{document}